# Novel Active Disturbance Rejection Control Based on Nested Linear Extended State Observers

Wameedh R. Abdul-Adheem, Ibraheem K. Ibraheem*, member, *IEEE*

*Abstract*—In this paper, a novel active disturbance rejection control (ADRC) strategy is proposed that replaces the Linear Extended state observer (LESO) used in conventional ADRC with a Nested LESO (N-LESO), functionally formed by connecting two LESOs in parallel. In the N-LESO, the inner LESO actively estimates and eliminates generalized disturbance in real-time; this includes the lumped exogenous disturbance, unwanted nonlinearities, and system uncertainties that a dynamical system generally exhibits. A closed form expression describing the error of the generalized disturbance to the bandwidth of the inner ESO has been derived. This states that, to reduce the generalized disturbance, it is necessary to increase the bandwidth of the inner ESO, which may conflict with H/W limitations and the sampling frequency of the system. Therefore, an alternative scenario is offered to reduce the remaining error of the generalized disturbance without increasing the bandwidth, provided that the rate of the change of the error is upper bounded. The stability of both LESO and N-ESO is investigated using Lyapunov stability analysis to show how the N-LESO is superior to the LESO in terms of reducing the generalized disturbance. Simulations on a second order uncertain nonlinear SISO system reveal that the proposed N-LESO can successfully deal with generalized disturbance in both noisy and noise-free environments.

*Keywords*—

## I. INTRODUCTION

The performance of a control system is excessively affected by system uncertainties, such as exogenous disturbances, unmodeled dynamics, and parameter perturbations. Guaranteeing simultaneously disturbance rejection and good tracking performance in light of the existence of large uncertainties complicates the design of any controller that aims to address these objectives. Accordingly, anti-disturbance methods with both external-loop controllers and internal-loop estimators have been comprehensively utilized. The precision of such controls mainly depends on the accuracy of the observer in the internal-loop. There have been various observer design philosophies posited, including fuzzy observers, sliding mode observers, unknown input observers, perturbation observers, equivalent input observers, extended state observers, and disturbance observers. Of these observers, the extended state observer (ESO) was originally suggested by Han [1]; it is often favoured because, in terms of design, it requires the minimum information from the system. It estimates the internal states of the system, system uncertainties, and exogenous disturbances, and it can also be used to design a state feedback controller. Based on this, an ESO is considered to be an essential part of the active disturbance rejection control paradigm. ESO-based control design has thus been widely examined in recent years [2]. The basic principle behind the operation of ESO is to augment the mathematical model of the nonlinear dynamical system with an additional virtual state that describes all the unwanted dynamics, uncertainties, and exogenous disturbances, which is termed "generalized disturbance" or "total disturbance". This virtual state, together with the states of the dynamic system, is observed in real-time using the ESO. This form of control design has been applied to a broad range of systems due to its model-independent operation. Initially, each ESO was constructed with nonlinear gains; however, it is more realistic to design and tune the ESO using tuneable linear gains, as proposed in [3]. Two signals, the input and the output of the nonlinear system, thus feed the ESO with information [4]. ESO-based control system design offers generally good performance due to the simplicity of design of ESO, which offers a need for minimum information, high precision of convergence, and fast-tracking capabilities [5]. In [6], ESO is tested on the nonlinear kinematic model of the differential drive mobile robot (DDMR). In [7], a general ESO-based control technique for non-chain integrator systems with mismatched disturbances was proposed. Recently, numerous control problems in various fields have also been effectively resolved by utilizing the ESO technique, including PMSM control [8], and attitude control of an aircraft [9]. The authors in [10] introduced an ESO-based dynamic sliding-mode control for high-order mismatched uncertainties with applications in motion control systems, and this also presented excellent tracking performance. In [11], an improved nonlinear ESO was proposed which achieved an outstanding performance in terms of smoothness in the control signal which leads to less control energy required to attain the desired performance.

In this paper, a novel ADRC is constructed by connecting a second ESO in parallel with an original ESO (the inner ESO), to construct a nested ADRC (N-ADRC). The advantage of this configuration is that the second ESO estimates and eliminates the remaining total disturbance that passes from the inner-ESO due to bandwidth limitations in real-time. Its excellent performance becomes very evident when considered in terms of measurement noise. To the best of the authors' knowledge, using double ESOs within the same ADRC structure, with applications in highly nonlinear uncertain systems, has not previously appeared in the literature.

An outline of this paper's contents and organisation follows. Section II briefly presents the concepts behind active disturbance rejection control (ADRC). A description of the proposed nested ESO and the relevant stability tests are included in section III. The numerical simulations verifying the validity of the proposed configuration are provided in section IV. Finally, the conclusion is given in section V, along with recommendations for future work.

## II. ACTIVE DISTURBANCE REJECTION CONTROL

In ADRC, the model of the nonlinear system is extended with an additional virtual state variable, which lumps all of the unwanted dynamics, uncertainties, and disturbances that remain unobserved in the standard system into a single term known as "generalized disturbance". In addition to estimating the states of the nonlinear system, the ESO performs online estimation and cancellation of this virtual state. In this scenario, the nonlinear system is converted into a chain of integrators, which allows the control system design to be simpler. Fig. 1 demonstrations the structure of a Conventional ADRC, (C-ADRC) which contains three key parts: the Tracking Differentiator (TD), an Extended State Observer (ESO), and Non-linear state error feedback (NLSEF) [12]. The tracking differentiator generates the required signal profile, which is the signal itself, free from noise, and a set of signal derivatives (1st derivative, 2nd derivative, …). The NLSEF acts as a nonlinear combination of the error profile. The ESO function is as discussed in the introduction section [13].

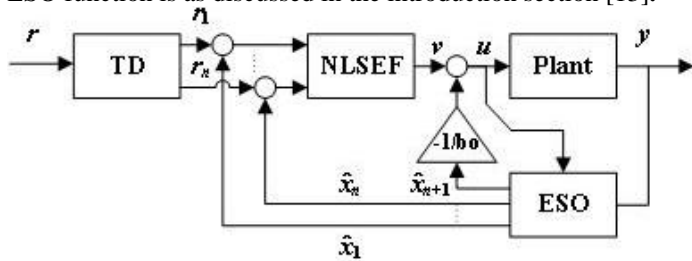

Fig. 1 Structure of ADRC.

### A. Tracking Differentiator (TD)

In the tracking differentiator, the output profile of the nonlinear system, in Brunovsky canonical form [14], must track the transient profile of the reference signal to resolve the problem of set-point jump in the traditional PID controller as stated in the seminal work [1]. In this manner, when a rapid change occurs in the set-point for any reason, the output signal of the plant will follow the output of the TD and will change gradually to reach the desired set-point [15]. TD can be represented as

$$\begin{cases} \dot{r}_1 = r_2 \\ \dot{r}_2 = -R\, sign(r_1 - r(t) + \frac{r_2 |r_2|}{2r}) \end{cases} \quad (1)$$

where $r_1$ is the tracking signal of the input $r$, and $r_2$ is the tracking signal of the derivative of the input $r$. To speed up or slow down the system during transient effects, the coefficient $R$ is adapted, making it application dependent [15]. Other versions of enhanced TD are proposed in [16]–[19].

### B. Non-Linear state error feedback

The linear weighting sum of the PID control is another limitation, involving as it does only the present, predictive, and accumulative errors, omitting other important parameters that could enhance its performance [15]. In the seminal work [1], the following nonlinear control law was suggested [1]:

$$fal(e.\alpha.\delta) = \begin{cases} \dfrac{e}{\delta^{1-\alpha}} & |x| \leq \delta \\ |e|^\alpha sign(e) & |x| \geq \delta \end{cases}$$

where α is a tuning parameter. The error signal, e, can thus reach zero more rapidly where α < 1 [15]. Other forms of nonlinear control laws are suggested in [20]–[23].

### C. ESO

Observers acquire data about the system states from its inputs and outputs progressively. Luenberger first recommended the rule of observers in [24], where it was concluded that the state vector of the system can be estimated by observing the input and output of the system. Subsequently, there have been numerous varieties of state observers outlined in the literature that rely upon the mathematical model of the system, including high gain observers and sliding mode observers [15]. The ESO was the first observer presented that was autonomous of the mathematical model and presented within the framework of ADRC. Furthermore, ESO has denoted estimators, which is considered a vital part of modern controls. The basic principle of the ESO is to observe the constituent parts of the generalized disturbance in real-time, including model discrepancy, exogenous disturbances, and the unmodeled dynamics of the nonlinear system. Additionally, it compensates for unpredicted disturbances in the control signal. ESO can be classified into two types. The first is linear ESO, which is an extension of the Luenberger observer [25], where the equations of the ESO contain only the linear correcting terms in order to simplify the calculations. These terms manipulate the error between the actual states of the system and the estimated states in such a way that the error approaches zero. The second type is nonlinear ESO, where the error correcting terms include a nonlinear function of the error. These nonlinear functions have the advantage of enhancing the estimation error more rapidly and smoothly than the linear ESO. Consider an *n-th* dimensional SISO nonlinear system:

$$\begin{cases} \dot{x}_1(t) = x_2(t) & x_1(0) = x_{10} \\ \dot{x}_2(t) = x_3(t) & x_2(0) = x_{20} \\ \quad \vdots \\ \dot{x}_n(t) = f(t, x_1(t), x_2(t), \dots, x_n(t)) + w(t) + u(t). & x_n(t) = x_{n0} \\ y(t) = x_1(t) \end{cases} \quad (1)$$

where $u(t) \in C(R,R)$ is the control input, $y(t)$ is the measured output, $f \in C(R_n, R)$ is an unknown system function, $w(t) \in C(R,R)$ is the uncertain exogenous disturbance, $x(t) = (x_1(t), x_2(t), \dots, x_n(t))$ is the state vector of the system, and $x(0) = (x_{10}, x_{20}, \dots, x_{n0})$ is the initial state. $L(t) = f + w(t)$ is therefore the "total disturbance" [1]. By adding the extended state $x_{n+1}(t) \stackrel{\text{def}}{=} f + w(t)$, system (1) can be written as:

$$\begin{cases} \dot{x}_1(t) = x_2(t) & x_1(0) = x_{10} \\ \dot{x}_2(t) = x_3(t) & x_2(0) = x_{20} \\ \quad \vdots \\ \dot{x}_n(t) = x_{n+1}(t) + u(t) & x(t) = x_{n0} \\ \dot{x}_{n+1}(t) = \dot{f}(t, x_1(t), x_2(t), \dots, x_n(t)) + \dot{w}(t) & x_{n+1}(t) = x_{n+1,0} \\ y(t) = x_1(t) \end{cases} \quad (2)$$

A linear extended state observer is given as per [1]:

$$\begin{cases} \dot{\hat{x}}_1(t) = \hat{x}_2(t) + \beta_1(y(t) - \hat{x}_1(t)) \\ \dot{\hat{x}}_2(t) = \hat{x}_3(t) + \beta_2(y(t) - \hat{x}_1(t)) \\ \vdots \\ \dot{\hat{x}}_n(t) = \hat{x}_{n+1}(t) + u(t) + \beta_n(y(t) - \hat{x}_1(t)) \\ \dot{\hat{x}}_{n+1}(t) = \beta_{n+1}(y(t) - \hat{x}_1(t)) \end{cases} \quad (3)$$

where $\beta_i$ is a constant observer gain to be tuned, $i = 1, 2, \ldots n + 1$, with $\beta_i = \frac{a_i}{\varepsilon^i}$.

Two approaches are common for ESO tuning: the pole-placement approach, and the bandwidth-based method. If the end goal is to reduce the number of parameters of the ESO, the parameters of the ESO can be expressed as a function of the bandwidth of the ESO, allowing only a single parameter of the ESO to be chosen or tuned. Selecting a bandwidth that is too large leads to a drop in the estimation error that nevertheless remains within an acceptable bound [26]. Observer bandwidth is chosen to be sufficiently larger than the disturbance frequency and smaller than the frequency of the unmodeled dynamics [27]. However, the performance of the ESO will deteriorate if the bandwidth of the ESO is selected to be too low or too high. High values in the bandwidth of the ESO and the controller result in good tracking performance and rejection of exogenous disturbances. The side effects of adopting large values for bandwidth can thus be summarized as 1) measurement noise causing a degradation in output tracking, introducing chatter on the control signal [28]; 2) a worsening of the transient response of the ESO, as large values of bandwidth lead to what are known as high gain observers [29]; and 3) the possibility of some unmodeled high-frequency dynamics being activated beyond a certain frequency, causing inconsistency in the closed-loop system. The noise and sampling rates are considered to be the two main factors constraining increases in the bandwidth. Based on this, an appropriate estimator bandwidth ought to be chosen in coordination with the noise tolerance and tracking performance. The authors in [30] designed a new class of adaptive ESO (AESO) in which the observer bandwidth varied with time to provide better performance than the LESO. The disadvantage of this method is that the parameter tuning may become more complex as AESO order increases [30]. To alleviate the peaking phenomenon caused by different initial values of the ESO, the small variable ε was designed as in [31]:

$$\frac{1}{\varepsilon} = \begin{cases} 100t^3 & 0 \le t \le 1 \\ 100 & t > 1 \end{cases}$$

The ESO parameters are tuned using Evolutionary Algorithm (EA) optimization techniques (BFO and PSO) rather than a manual process. Eventually, the ESO begins estimating these states. Consequently, the effect of lumped disturbances is canceled and the controller actively compensates for the disturbances in real time [25].

## III. MAIN RESULTS

The innovative ADRC is constructed by adding an extra LESO, which shares an output signal with the plant to be controlled with inner loop LESO. The structure of the novel ADRC is presented in Fig. 2. The inner LESO accomplishes the estimation of plant states and total disturbance. In a situation where a suitably low bandwidth frequency is selected for the inner LESO to reduce noise, the estimation of the generalized disturbance through the augmented state is associated with a relatively large estimation error. The outer loop LESO will thus complete the rejection process by choosing an appropriate control law.

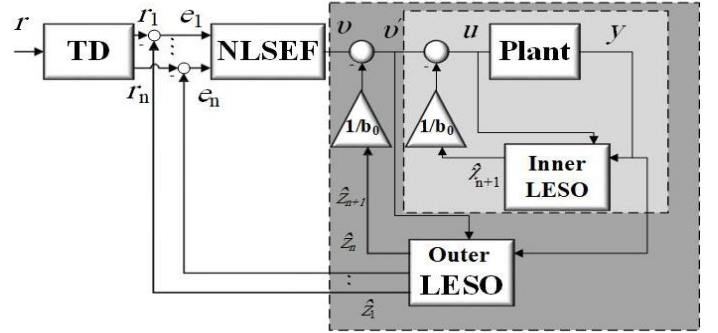

Fig. 2 The novel ADRC(N-ADRC) structure with nested LESOs.

**Assumption 1:** The function $L$ is continuously differentiable with a local Lipschitz derivative.

**Assumption 2:** There is a positive constant $M$ such that $|\Delta(t)| \le M$ for $t \ge 0$.

**Assumption 3:** There exist constants $\lambda_1$ and $\lambda_2$ and positive definite, continuous differentiable functions $V, W: \mathbb{R}^{n+1} \to \mathbb{R}^+$ such that

$$\lambda_1 \|y\|^2 \le V(y) \le \lambda_2 \|y\|^2 \, . \, W(y) = \|y\|^2; \quad (4)$$

$$\sum_{i=1}^{\rho} \frac{\partial V_i}{\partial y_i}(y_i - a_i y_1) - \frac{\partial V}{\partial y_{\rho+1}} a_{\rho+1} y_1 \le -W(y). \quad (5)$$

**Lemma 1.** Consider the candidate Lyapunov functions $V, W: \mathbb{R}^{n+1} \to \mathbb{R}^+$ defined by $V(\eta) = <P\eta, \eta> = \eta^T P \eta$, where $\eta \in \mathbb{R}^{\rho+1}$ and $P$ is a symmetric and positive definite matrix. Suppose assumption 3 (7) with $\lambda_1 = \lambda_{min}(P)$ and $\lambda_2 = \lambda_{max}(P)$, where $\lambda_{min}(P)$ and $\lambda_{max}(P)$ are the minimal and maximal eigenvalues of $P$, respectively. Then, (i) $\left|\frac{\partial V}{\partial \eta_{\rho+1}}\right| \le 2\lambda_{max}(P)\|\eta\|$, (ii) $-W(\eta) \le \frac{V(\eta)}{\lambda_{max}(P)}$, and (iii) $\|\eta\| \le \sqrt{\frac{V(\eta)}{\lambda_{min}(P)}}$.

**Proof:** As $V(\eta) \le \lambda_{max}(P)\|\eta\|^2$ and $\left|\frac{\partial V}{\partial \eta_{n+1}}\right| \le \left\|\frac{\partial V(\eta)}{\partial \eta}\right\|$, then $\left|\frac{\partial V}{\partial \eta_{\rho+1}}\right| \le 2\lambda_{max}(P)\|\eta\|$. As $V(\eta) \le \lambda_{max}(P)\|\eta\|^2 = \lambda_{max}(P)W(\eta)$. Thus, $-W(\eta) \le \frac{V(\eta)}{\lambda_{max}(P)}$. Finally, because $\lambda_{min}(P)\|\eta\|^2 \le V(\eta)$, this leads to $\|\eta\| \le \sqrt{\frac{V(\eta)}{\lambda_{min}(P)}}$.

**Theorem 1.** Given the nonlinear plant (2) and the linear extended state observer (3),

$$\lim_{t \to \infty} |x_i(t) - \hat{x}_i(t)| = \frac{1}{\omega_0^{n+2-i}} \frac{2M\lambda_{max}^2(P)}{\lambda_{min}(P)}$$

where $x_i(t)$, and $\hat{x}_i(t)$ denote the solutions of (2) and (3) respectively, $i \in \{1,2,...,n+1\}$, and $\beta_i = a_i\omega_0^i$; $a_i$, and $i \in \{1,2,...n+1\}$ are relevant constants, and $\omega_0$ is the bandwidth constant.

**Proof:** Based on the work in [1], the proof is as follows:
Set $e_i(t) = x_i(t) - \hat{x}_i(t)$, for $i \in \{1,2,,n+1\}$. (6)
Subtracting (3) from (2) gives

$$\begin{cases} \dot{x}_1(t) - \dot{\hat{x}}_1(t) = x_2(t) - (\hat{x}_2(t) + \beta_1(y(t) - \hat{x}_1(t))) \\ \dot{x}_2(t) - \dot{\hat{x}}_2(t) = x_3(t) - (\hat{x}_3(t) + \beta_2(y(t) - \hat{x}_1(t))) \\ \vdots \\ \dot{x}_n(t) - \dot{\hat{x}}_n(t) = x_{n+1}(t) + u(t) - (\hat{x}_{n+1}(t) + u(t) + \beta_n(y(t) - \hat{x}_1(t))) \\ \dot{x}_{n+1}(t) - \dot{\hat{x}}_{n+1}(t) = \Delta(t) - \beta_{n+1}(y(t) - \hat{x}_1(t)). \end{cases}$$

Direct computation shows that the estimation error dynamics satisfy

$$\begin{cases} \dot{e}_1(t) = e_2(t) - \beta_1 e(t) \\ \dot{e}_2(t) = e_3(t) - \beta_2 e(t) \\ \vdots \\ \dot{e}_n(t) = e_{n+1}(t) - \beta_n e(t) \\ \dot{e}_{n+1}(t) = \Delta(t) - \beta_{n+1} e(t), \end{cases} \quad (7)$$

and thus the final form is

$$\begin{cases} \dot{e}_1(t) = e_2(t) - \omega_0 a_1 \cdot e_1(t) \\ \dot{e}_2(t) = e_3(t) - \omega_0^2 a_2 e_1(t) \\ \vdots \\ \dot{e}_n(t) = e_{i,n+1}(t) - \omega_0^n a_n \cdot e_1(t) \\ \dot{e}_{n+1}(t) = \Delta(t) - \omega_0^{n+1} a_{n+1} \cdot e_1(t) \end{cases} \quad (8)$$

Set

$$\eta_i(t) = \omega_0^{n+1-i} e_i\left(\frac{t}{\omega_{i0}}\right), i \in \{1,2,,n+1\} \quad (9)$$

or $e_i\left(\frac{t}{\omega_0}\right) = \frac{1}{\omega_0^{n+1-i}} \eta_i(t)$

$$\begin{cases} \frac{de_1\left(\frac{t}{\omega_0}\right)}{d\frac{t}{\omega_0}} = e_2\left(\frac{t}{\omega_0}\right) - \omega_0 a_1 \cdot e_1\left(\frac{t}{\omega_0}\right) \\ \frac{de_2\left(\frac{t}{\omega_{i0}}\right)}{d\frac{t}{\omega_0}} = e_3\left(\frac{t}{\omega_0}\right) - \omega_0^2 a_2 \cdot e_1\left(\frac{t}{\omega_0}\right) \\ \vdots \\ \frac{de_n\left(\frac{t}{\omega_{i0}}\right)}{d\frac{t}{\omega_0}} = e_{n+1}\left(\frac{t}{\omega_0}\right) - \omega_0^n a_n \cdot e_1\left(\frac{t}{\omega_0}\right) \\ \frac{de_{n+1}\left(\frac{t}{\omega_{i0}}\right)}{d\frac{t}{\omega_0}} = \Delta - \omega_0^{n+1} a_{n+1} \cdot e_1\left(\frac{t}{\omega_0}\right) \end{cases} \quad (10)$$

From (9), $\frac{d\eta_i(t)}{dt} = \omega_0^{n+1-i} \frac{de_i\left(\frac{t}{\omega_0}\right)}{d\frac{t}{\omega_0}} \frac{d\left(\frac{t}{\omega_0}\right)}{dt} = \omega_{i0}^{n-i} \frac{de_i\left(\frac{t}{\omega_0}\right)}{d\frac{t}{\omega_0}}$ can be derived. Then,

$$\frac{de_i\left(\frac{t}{\omega_0}\right)}{d\frac{t}{\omega_0}} = \frac{1}{\omega_0^{n-i}} \frac{d\eta_i(t)}{dt} \quad . \quad (11)$$

Both (9) and (11) are substituted into (10) and the result is

$$\begin{cases} \frac{1}{\omega_0^{n-1}} \frac{d\eta_1(t)}{dt} = \frac{1}{\omega_0^{n-1}} \eta_2(t) - \omega_0 a_1 \cdot \frac{1}{\omega_0^n} \eta_1(t) \\ \frac{1}{\omega_0^{n-2}} \frac{d\eta_2(t)}{dt} = \frac{1}{\omega_0^{n-2}} \eta_3(t) - \omega_0^2 a_2 \cdot \frac{1}{\omega_0^n} \eta_1(t) \\ \vdots \\ \frac{d\eta_n(t)}{dt} = \eta_{in+1}(t) - \omega_0^n a_n \cdot \frac{1}{\omega_0^n} \eta_i(t) \\ \frac{1}{\omega_0^{-1}} \frac{d\eta_{i,n+2}(t)}{dt} = \Delta - \omega_0^{n+1} a_{n+1} \cdot \frac{1}{\omega_0^n} \eta_1(t) \end{cases} \quad .$$

The time-scaled estimation error dynamics are

$$\begin{cases} \frac{d\eta_1(t)}{dt} = \eta_2(t) - a_1 \eta_1(t) \\ \frac{d\eta_2(t)}{dt} = \eta_3(t) - a_2 \eta_1(t) \\ \vdots \\ \frac{d\eta_n(t)}{dt} = \eta_{n+1}(t) - a_n \eta_i(t) \\ \frac{d\eta_{n+1}(t)}{dt} = \frac{\Delta}{\omega_0} - a_{n+1} \cdot \eta_1(t) \end{cases} \quad .$$

(12)

Finding the derivative of $V(\eta)$ (from lemma 1) with respect to $t$ along solution $\eta$ along (12), $\dot{V}(\eta)|_{along\ (12)} = \sum_{i=1}^{n+1} \frac{\partial V(\eta)}{\partial \eta_i} \dot{\eta}_i(t) = \sum_{i=1}^{n+1} \frac{\partial V(\eta)}{\eta_i} (\eta_{i+1}(t) - a_i \cdot \eta_1(t)) + \frac{\partial V(\eta)}{\partial \eta_{n+1}} \left(\frac{\Delta}{\omega_0} - a_{n+1} \cdot \eta_1(t)\right)$.

Then, $\dot{V}(\eta)|_{along\ (12)} = \sum_{i=1}^{n+1} \frac{\partial V(\eta)}{\eta_i} (\eta_{i+1}(t) - a_i \cdot \eta_1(t)) + \frac{\partial V(\eta)}{\partial \eta_{n+1}} \frac{\Delta}{\omega_0} - \frac{\partial V(\eta)}{\partial \eta_{n+1}} a_{n+1} \cdot \eta_1(t)$.

From assumption 2, $\dot{V}(\eta)|_{along\ (12)} \leq -W(\eta) + \frac{\partial V(\eta)}{\partial \eta_{n+1}} \frac{\Delta}{\omega_0}$. From assumption 1 and the results of Lemma 1, $\dot{V}_i(\eta_i) \leq -\frac{V(\eta)}{\lambda_{max}(P)} + \frac{M}{\omega_0} 2\lambda_{max}(P) \frac{\sqrt{V_i(\eta)}}{\sqrt{\lambda_{min}(P)}}$. As $\frac{d}{dt}\sqrt{V(\eta)} = \frac{1}{2}\frac{1}{\sqrt{V(\eta)}} \dot{V}(\eta)$, $\frac{d}{dt}\sqrt{V(\eta)} \leq \frac{1}{2}\frac{1}{\sqrt{V(\eta)}}\left(-\frac{V(\eta)}{\lambda_{max}(P)} + \frac{M}{\omega_0} 2\lambda_{max}(P) \frac{\sqrt{V(\eta)}}{\sqrt{\lambda_{min}(\eta)}}\right)$.

Thus, $\frac{d}{dt}\sqrt{V(\eta)} \leq -\frac{\sqrt{V(\eta)}}{2\lambda_{max}(P)} + \frac{M}{\omega_0} \frac{\lambda_{max}(P)}{\sqrt{\lambda_{min}(P)}}$ . (13)

Solving ordinary differential equation (13) gives

$$\sqrt{V(\eta)} \leq \frac{2M\lambda_{max}^2(P)}{\omega_0\sqrt{\lambda_{min}(P)}}\left(1 - e^{-\frac{t}{2\lambda_{max}(P)}}\right) + \sqrt{V(\eta(0))} e^{-\frac{t}{2\lambda_{max}(P)}}$$

$$\|\eta(t)\| \leq \sqrt{\frac{1}{\lambda_{min}(P)}}\left(\frac{2M\lambda_{max}^2(P)}{\omega_0\sqrt{\lambda_{min}(P)}}\left(1 - e^{-\frac{t}{2\lambda_{max}(P)}}\right) + \sqrt{V(\eta(0))} e^{-\frac{t}{2\lambda_{max}(P)}}\right)$$

$$\|\eta(t)\| \leq \frac{2M\lambda_{max}^2(P)}{\omega_0\lambda_{min}(P)}\left(1 - e^{-\frac{t}{2\lambda_{max}(P)}}\right) + \sqrt{\frac{V(\eta(0))}{\lambda_{min}(P)}} e^{-\frac{t}{2\lambda_{max}(P)}} \quad (14)$$

It follows from (9) that $|x_i(t) - \hat{x}_i(t)| = \frac{1}{\omega_{i0}^{n+1-i}}|\eta_i(\omega_0 t)| \Rightarrow$
$|x_i(t) - \hat{x}_i(t)| \leq \frac{1}{\omega_0^{n+1-i}}\|\eta(t)\|$. Thus, by using (13),
$|x_i(t) - \hat{x}_i(t)| \leq \frac{1}{\omega_0^{n+1-i}}\left(\frac{2M\lambda_{max}^2(P)}{\omega_0 \lambda_{min}(P)}\left(1 - e^{-\frac{t}{2\lambda_{max}(P)}}\right) + \sqrt{\frac{V(\eta(0))}{\lambda_{min}(P)}} e^{-\frac{t}{2\lambda_{max}(P)}}\right)$
Finally,
$$\lim_{t \to \infty} |x_i(t) - \hat{x}_i(t)| \leq \frac{1}{\omega_0^{n+2-i}} \frac{2M\lambda_{max}^2(P)}{\lambda_{min}(P)}. \tag{15}$$
From (6), setting $i=n+1$ gives
$$e_{n+1} = x_{n+1} - \hat{x}_{n+1} \Rightarrow x_{n+1} = e_{n+1} + \hat{x}_{n+1} \tag{16}$$
Consider the control law described by
$$u(t) = v'(t) - \hat{x}_{n+1} \quad . \tag{17}$$
Substituting (16) and (17) into (2) gives
$$\begin{cases} \dot{x}_1(t) = x_2(t) \\ \dot{x}_2(t) = x_3(t) \\ \vdots \\ \dot{x}_n(t) = e_{n+1} + v'(t) \\ y(t) = x_1(t) \end{cases} \tag{18}$$
Adding an augmented state to the resultant system (18) thus creates
$$\begin{cases} \dot{x}_1(t) = x_2(t) \\ \dot{x}_2(t) = x_3(t) \\ \vdots \\ \dot{x}_n(t) = x_{n+1} + v(t) \\ \dot{x}_{n+1} = \Delta' = \dot{e}_{n+1} \\ y(t) = x_1(t) \end{cases} \tag{19}$$
and the outer LESO can be described by
$$\begin{aligned} \dot{\hat{z}}_1(t) &= \hat{z}_2(t) + l_1(y(t) - \hat{z}_1(t)) \\ \dot{\hat{z}}_2(t) &= \hat{z}_3(t) + l_2(y(t) - \hat{z}_1(t)) \\ &\vdots \\ \dot{\hat{z}}_n(t) &= \hat{z}_{n+1}(t) + v(t) + l_n(y(t) - \hat{z}_1(t)) \\ \dot{\hat{z}}_{n+1}(t) &= l_{n+1}(y(t) - \hat{z}_1(t)) \end{aligned} \tag{20}$$

**Lemma 2.** Consider the system given in (2), and the linear extended state observer (3). The upper bound of the derivative of the generalized disturbance estimation error is given by $\lim_{\substack{t \to \infty \\ a_{n+1} \to 0}} |\dot{e}_{n+1}| \leq M'$, where $M' \leq M$.

**Proof:** From (6), with $i = n + 1$, $e_{n+1} = x_{n+1} - \hat{x}_{n+1} \Rightarrow \dot{e}_{n+1} = \dot{x}_{n+1} - \dot{\hat{x}}_{n+1}$. Thus, $|\dot{e}_{n+1}| \leq |\dot{x}_{n+1}| + |\dot{\hat{x}}_{n+1}|$, and from (2) and (3),
$$|\dot{e}_{n+1}| \leq |\Delta(t)| + |\beta_{n+1} e_1(t)| \tag{21}$$
From (15), $\lim_{t \to \infty} |e_1(t)| \leq \frac{1}{\omega_0^{n+1}} \frac{2M\lambda_{max}^2(P)}{\lambda_{min}(P)}$. As $\beta_{n+1} = a_{n+1}\omega_0^{n+1}$, $\lim_{t \to \infty} |\beta_{n+1} e_1(t)| \leq a_{n+1} \frac{2M\lambda_{max}^2(P)}{\lambda_{min}(P)}$.

Thus, $\lim_{\substack{t \to \infty \\ a_{n+1} \to 0}} |\beta_{n+1} e_1(t)| = 0 \tag{22}$

From (21) and (22), $\lim_{\substack{t \to \infty \\ a_{n+1} \to 0}} |\dot{e}_{n+1}| \leq |\Delta(t)|$, and $\lim_{\substack{t \to \infty \\ a_{n+1} \to 0}} |\dot{e}_{n+1}| \leq M$. Consider $M'$ such that

$$\lim_{\substack{t \to \infty \\ a_{n+1} \to 0}} |\dot{e}_{n+1}| \leq M' \leq M \tag{23}$$

**Corollary 1.** Consider the system given in (19), and the linear extended state observer (20). Here, $\lim_{t \to \infty} |x_i(t) - \hat{z}_i(t)| \leq \frac{1}{\omega_0'^{n+2-i}} \frac{2M'\lambda_{max}^2(P')}{\lambda_{min}(P')}$, where $x_i(t)$, and $\hat{z}_i(t)$ denote the solutions to (19) and (20) respectively, $i \in \{1,2,\ldots,n+1\}$, and $l_i = \alpha_i \omega_0'^{i}$, where $\alpha_i, i \in \{1,2,\ldots,n+1\}$ are relevant constants, and $\omega_0'$ is the bandwidth constant of the outer LESO.

**Proof:** as in Theorem 1, let
$$\zeta_i(t) = x_i(t) - \hat{z}_i(t), i \in \{1,2,\ldots,n+1\}, \tag{24}$$
and $\gamma_i(t) = \omega_0'^{n+1-i} \xi_i(\frac{t}{\omega_0'})$, $i \in \{1,2,\ldots,n+1\}$.

Thus, $|x_i(t) - \hat{z}_i(t)| \leq \frac{1}{\omega_0'^{n+1-i}}\left(\frac{2M'\lambda_{max}^2(P')}{\omega_0' \lambda_{min}(P')}\left(1 - e^{-\frac{t}{2\lambda_{max}(P')}}\right) + \sqrt{\frac{V(\gamma(0))}{\lambda_{min}(P')}} e^{-\frac{t}{2\lambda_{max}(P')}}\right)$ and

$$\lim_{t \to \infty} |x_i(t) - \hat{z}_i(t)| \leq \frac{1}{\omega_0'^{n+2-i}} \frac{2M'\lambda_{max}^2(P')}{\lambda_{min}(P')} \quad . \tag{25}$$

## IV. NUMERICAL SIMULATION

The classical ADRC, given in fig.1 was first implemented to reject the total disturbances from the theoretical system described in (26):
$$\begin{cases} \dot{x}_1 = x_2 \\ \dot{x}_2 = f(x_1, x_2) + w(t) + (1 + a_3 \sin(t))u \\ y = x_1 \end{cases} \tag{26}$$
where the unknown function is $f(x_1, x_2) = a_1 x_1 + a_2 \sin(x_2)$, given that $a_1 = 0.2$, $a_2 = 0.1$, and the exogenous disturbance $w(t) = \exp(-t)\cos(t)$. After this, the novel ADRC based on N-LESO was also implemented for system (26). Both controllers and the suggested system were numerically simulated using MATLAB®/Simulink® ODE45 solver for models with continuous states. The reference input to the system was $\cos(0.5t)$ applied at t = 0 sec. Two test conditions were considered for this work. In the first case, the output of the proposed system did not include any measurement noise, while in the second test case, a Gaussian noise was applied with variance equal to $10^{-4}$ and zero mean. The simulation results of both conventional ADRC and N-ADRC are shown in fig. 3. The numerical results are listed in table I. Adding measurement noise to the measured output significantly affected the output response (ITAE) and the total energy of the actuating signal (ISU) of the C-ADRC controller. The N-ADRC shows a noticeable reduction in both ITAE and ISU.

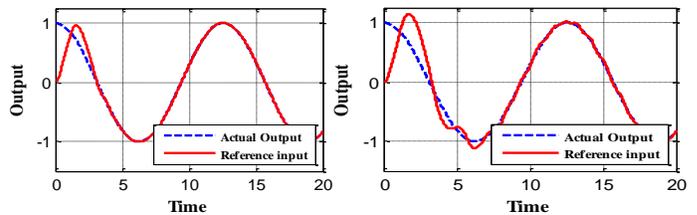

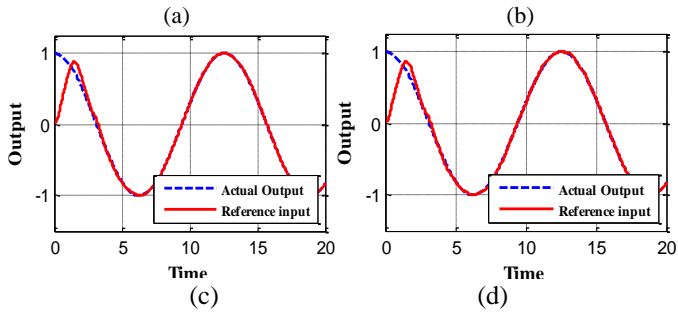

Fig.3. The response curves.
(a) C-ADRC (without noise) (b) C-ADRC (with Gaussian noise)
(c) N-ADRC (without noise) (d) N- ADRC (with Gaussian noise)

Table I Numerical results

|  | Without noise | | With noise | |
| --- | --- | --- | --- | --- |
|  | ITAE | ISU | ITAE | ISU |
| C-ADRC | 1.714331 | 7.168854 | 7.066944 | 457.302382 |
| N-ADRC | 1.331774 | 6.630304 | 2.128657 | 310.913988 |
| Reduction(%) | 22.32 | 7.51 | 69.87 | 32.01 |

The estimation error of the total disturbances for the inner LESO is described by $e_3$ which is given in (6), and the total disturbance estimation error of the outer LESO is described by $\zeta_3$ which is given in (24); both of these are illustrated in fig. 4. The ITAE of $e_3$ is 10.5769 and the ITAE of $\zeta_3$ is 5.8251, displaying a percentage reduction in the ITAE equal to 45%. Fig. 4 more clearly illustrates the reduction in $\zeta_3$ against $e_3$.

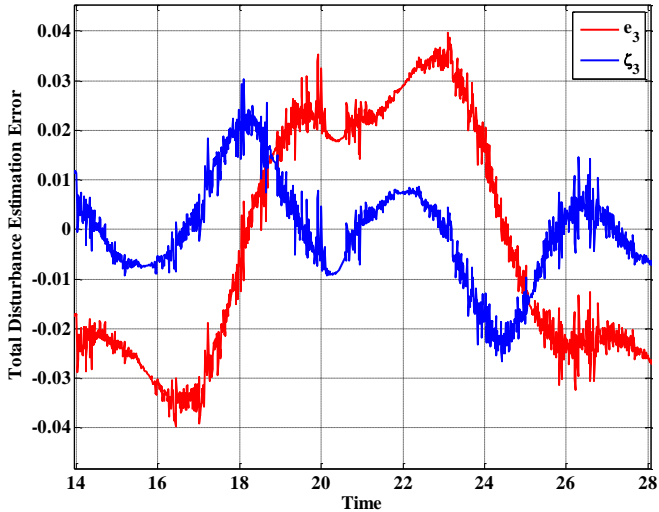

Fig.4 Total disturbance estimation errors.

## V. CONCLUSIONS

This paper presented a novel approach to the design of a new class of ESO achieved by nesting an additional ESO in parallel with the original to obtain an N-ADRC. The proposed N-ADRC was successfully applied to the hypothetical SISO and a highly nonlinear system with exogenous disturbance as given in (26). It can be concluded that the N-ADRC outperforms the C-ADRC in terms of control effort, output tracking, and disturbance rejection, as well as, more obviously, in the case of measurement error. In contrast with the C-ADRC, where increasing bandwidth is the only option for obtaining better performance, the main outcome of this work is to show that a second ESO connected in parallel with the internal ESO removes the need to increase the bandwidth of the internal ESO. Furthermore, the N-ADRC can converge to the states of the original system asymptotically. The N-ADRC reduced the ITAE dramatically for cases both with and without measurement noise. in future work, this approach can be extended to nest more than two LESOs, and nonlinear ESOs could also be used and their performance investigated.

REFERENCES

[1] J. Han, "From PID to active disturbance rejection control," *IEEE Trans. Ind. Electron.*, vol. 56, no. 3, pp. 900–906, 2009.
[2] Y. Zhang, J. Zhang, L. Wang, and J. Su, "Composite disturbance rejection control based on generalized extended state observer," *ISA Trans.*, vol. 63, pp. 377–386, 2016.
[3] Z. Gao, Y. Huang, and J. Han, "An alternative paradigm for control system design," *Proc. IEEE Conf. Decis. Control*, vol. 5, pp. 4578–4585, 2001.
[4] A. H. M. Sayem, Z. Cao, and Z. Man, "Model Free ESO-based Repetitive Control for Rejecting Periodic and Aperiodic Disturbances," *IEEE Trans. Ind. Electron.*, vol. PP, no. 99, pp. 1–8, 2016.
[5] G. Li, W. Xu, J. Zhao, S. Wang, and B. Li, "Precise robust adaptive dynamic surface control of permanent magnet synchronous motor based on extended state observer," *IET Sci. Meas. Technol.*, vol. 11, no. 5, pp. 590–599, 2017.
[6] I. K. Ibraheem and W. R. Abdul-Adheem, "An Improved Active Disturbance Rejection Control for a Differential Drive Mobile Robot with Mismatched Disturbances and Uncertainties," pp. 7–12, 2017.
[7] S. Li, J. Yang, W. H. Chen, and X. Chen, "Generalized extended state observer based control for systems with mismatched uncertainties," *IEEE Trans. Ind. Electron.*, vol. 59, no. 12, pp. 4792–4802, 2012.
[8] H. Liu and S. Li, "Speed control for PMSM servo system using predictive functional control and extended state observer," *IEEE Trans. Ind. Electron.*, vol. 59, no. 2, pp. 1171–1183, 2012.
[9] Z. Zhu, Y. Xia, and M. Fu, "Adaptive sliding mode control for attitude stabilization with actuator saturation," *IEEE Trans. Ind. Electron.*, vol. 58, no. 10, pp. 4898–4907, 2011.
[10] J. Yang, J. Su, S. Li, and X. Yu, "High-order mismatched disturbance compensation for motion control systems via a continuous dynamic sliding-mode approach," *IEEE Trans. Ind. Informatics*, vol. 10, no. 1, pp. 604–614, 2014.
[11] W. R. Abdul-adheem and I. K. Ibraheem, "Improved Sliding Mode Nonlinear Extended State Observer based Active Disturbance Rejection Control for Uncertain Systems with Unknown Total Disturbance," *Int. J. Adv. Comput. Sci. Appl.*, vol. 7, no. 12, pp. 80–93, 2016.
[12] P. Jiang, J. Y. Hao, X. P. Zong, and P. G. Wang, "Modeling and simulation of Active-Disturbance-Rejection Controller with Simulink," *2010 Int. Conf. Mach. Learn. Cybern. ICMLC 2010*, vol. 2, no. July, pp. 927–931, 2010.
[13] R. Parvathy and A. E. Daniel, "A survey on active disturbance rejection control," *2013 Int. Mutli-Conference Autom. Comput. Commun. Control Compress. Sens.*, no. 4, pp. 330–335, 2013.
[14] M. Nowicki, R. Madoński, and K. Kozłowski, "First look at conditions on applicability of ADRC," *2015 10th Int. Work. Robot Motion Control. RoMoCo 2015*, pp. 294–299, 2015.
[15] F. Al-Kalbani, S. M. Al Hosni, and J. Zhang, "Active Disturbance Rejection Control of a methanol-water separation distillation column,"

*2015 IEEE 8th GCC Conf. Exhib.*, no. March, pp. 1–6, 2015.

[16] H. Lin and X. Wang, "Design and analysis of a continuous hybrid differentiator," *IET Control Theory Appl.*, vol. 5, no. 11, pp. 1321–1334, 2011.

[17] X. Wang and B. Shirinzadeh, "Rapid-convergent nonlinear differentiator," *Mech. Syst. Signal Process.*, vol. 28, pp. 414–431, 2012.

[18] M. T. Angulo, J. A. Moreno, and L. Fridman, "Robust exact uniformly convergent arbitrary order differentiator," *Automatica*, vol. 49, no. 8, pp. 2489–2495, 2013.

[19] I. K. Ibraheem and W. R. Abdul-adheem, "On the Improved Nonlinear Tracking Differentiator based Nonlinear PID Controller Design," vol. 7, no. 10, pp. 234–241, 2016.

[20] W. R. Abdul-adheem, "From PID to Nonlinear State Error Feedback Controller," vol. 8, no. 1, pp. 312–322, 2017.

[21] G. B. S. Y.L.Kang and T.T.Lie, "Application of an NLPID controller on a UPFC to improve transient stability of a power system," vol. 148, no. 6, 2001.

[22] L. Ma, F. Lin, X. You, and T. Q. Zheng, "Nonlinear PID Control of Three-Phase Pulse Width Modulation Rectifier," no. 2, pp. 3417–3422, 2008.

[23] S. N. S. Salim, Z. H. Ismail, M. F. Rahmat, a. a. M. Faudzi, N. H. Sunar, and S. I. Samsudin, "Tracking performance and disturbance rejection of pneumatic actuator system," *2013 9th Asian Control Conf.*, pp. 1–6, 2013.

[24] D. G. Luenberger, "Observing the State of a Linear System," *IEEE Trans. Mil. Electron.*, vol. 8, no. 2, pp. 74–80, 1964.

[25] A. Goel and A. Swarup, "Performance Analysis of Active Disturbance Rejection Controlled Robotic Manipulator based on Evolutionary Algorithm," *Int. J. Hybrid Inf. Technol.*, vol. 9, no. 1, pp. 65–80, 2016.

[26] D. Bao and W. Tang, "Adaptive sliding mode control of ball screw drive system with extended state observer," *Proc. - 2016 2nd Int. Conf. Control. Autom. Robot. ICCAR 2016*, no. 2, pp. 133–138, 2016.

[27] A. A. Godbole, J. P. Kolhe, and S. E. Talole, "Performance analysis of generalized extended state observer in tackling sinusoidal disturbances," *IEEE Trans. Control Syst. Technol.*, vol. 21, no. 6, pp. 2212–2223, 2013.

[28] H. Pan, W. Sun, H. Gao, T. Hayat, and F. Alsaadi, "Nonlinear tracking control based on extended state observer for vehicle active suspensions with performance constraints," *Mechatronics*, vol. 30, pp. 363–370, 2015.

[29] B. Z. Guo and Z. L. Zhao, "On the convergence of an extended state observer for nonlinear systems with uncertainty," *Syst. Control Lett.*, vol. 60, no. 6, pp. 420–430, 2011.

[30] Z. Pu, R. Yuan, J. Yi, and X. Tan, "A Class of Adaptive Extended State Observers for Nonlinear Disturbed Systems," *IEEE Trans. Ind. Electron.*, vol. 62, no. 9, pp. 5858–5869, 2015.

[31] Y. Li, B. Yang, T. Zheng, Y. Li, M. Cui, and S. Peeta, "Extended-State-Observer-Based Double-Loop Integral Sliding-Mode Control of Electronic Throttle Valve," *IEEE Trans. Intell. Transp. Syst.*, vol. 16, no. 5, pp. 2501–2510, 2015.